\documentclass[a4paper,11pt]{article}
\pdfoutput=1 

\usepackage{jinstpub} 
\usepackage[normalem]{ulem}

\title{SuperFGD prototype time resolution studies}

\author[a]{I.~Alekseev,}
\author[b]{T.~Arihara,}

\author[c]{V.~Baranov,}
\author[d]{L.~Bartoszek,}
\author[e]{L.~Bernardi,}
\author[f,g]{A.~Blondel,}

\author[c]{A.V.~Boikov,}
\author[e]{M.~Buizza-Avanzini,}

\author[f]{F. Cadoux,}
\author[h]{J.~Cap\'o,}
\author[i]{J.~Cayo,}
\author[e]{J.~Chakrani,}
\author[i]{P.S.~Chong,}
\author[j]{A.~Chvirova,}

\author[a]{M.~Danilov,}
\author[c]{Yu.I.~Davydov,}
\author[j]{A.~Dergacheva,}
\author[k]{N.~Dokania,}
\author[f]{D.~Douqa,}
\author[e]{O.~Drapier,}

\author[l]{A.~Eguchi,}

\author[f]{Y.~Favre,}
\author[j]{D.~Fedorova,}
\author[j]{S.~Fedotov,}
\author[m]{Y.~Fujii,}

\author[e]{F.~Gastaldi,}
\author[n]{A.~Gendotti,}
\author[c]{V.~Glagolev,}
\author[e]{R.~Guillaumat,}

\author[l]{K.~Iwamoto,}

\author[m]{M.~Jakkapu,}
\author[h]{C.~Jes\'us-Valls,}
\author[k]{C.K.~Jung,}

\author[b]{H. Kakuno,}
\author[o]{S.P.~Kasetti,}
\author[j]{M.~Khabibullin,}
\author[j]{A.~Khotjantsev,}
\author[l]{H.~Kikutani,}
\author[m]{T.~Kobayashi,}
\author[l]{S.~Kodama,}
\author[f]{A.~Korzenev,}
\author[n]{U.~Kose,}
\author[j,p,q]{Y.~Kudenko,}
\author[o]{T.~Kutter,}

\author[i]{D.~Last,}
\author[n]{B.~Li,}
\author[i]{Z.~Li,}
\author[i]{L.S.~Lin,}
\author[o]{S.~Lin,}
\author[e]{M.~Louzir,}
\author[h]{T.~Lux,}

\author[f]{L.~Maret,}
\author[k]{S.~Martynenko,}
\author[m]{T.~Matsubara,}
\author[i]{C.~Mauger,}
\author[k]{C.~McGrew,}
\author[j]{A.~Mefodiev,}
\author[j]{O.~Mineev,}

\author[m]{T.~Nakadaira,}
\author[l]{K.~Nakagiri,}
\author[e]{J.~Nanni,}
\author[f]{L.~Nicola,}
\author[f]{E.~Noah,}

\author[r]{V.~Paolone,}
\author[f]{S.~Parsa,}
\author[i]{R.~Pellegrino,}

\author[i]{M.A.~Ramirez,}
\author[d]{M.~Reh}
\author[k]{C.~Ricco,}
\author[n]{A.~Rubbia,}

\author[m]{K.~Sakashita,}
\author[f]{F.~Sanchez,}
\author[n]{D.~Sgalaberna,}
\author[j]{A.~Shvartsman,}
\author[a]{N.~Skrobova,\note{Corresponding author.}}
\author[c]{I.A.~Suslov,}
\author[j,r]{S.~Suvorov,}
\author[a]{D.~Svirida,}

\author[k]{A.~Teklu,}
\author[c]{V.V.~Tereshchenko,}
\author[o]{M.~Tzanov,}

\author[c]{I.I.~Vasilyev,}

\author[k]{K.~Wood,}

\author[k]{G.~Yang,}
\author[j]{N.Yershov,}
\author[l]{M.~Yokoyama,}
\author[l]{Y.~Yoshimoto,}

\author[n]{X.~Zhao,}
\author[k]{P.~Zilberman,}
\author[d]{E.~D.~Zimmerman}

\affiliation[a]{Lebedev Physical Institute of the Russian Academy of Sciences, 53 Leninskiy Prospekt, Moscow, 119991, Russia}
\affiliation[b]{Tokyo Metropolitan University, Department of Physics, Tokyo, Japan}
\affiliation[c]{Joint Institute for Nuclear Research, Dubna, Moscow Region, Russia}
\affiliation[d]{University of Colorado, Boulder, Colorado 80309 USA}
\affiliation[e]{Ecole Polytechnique, IN2P3-CNRS, Laboratoire Leprince-Ringuet, Palaiseau, France}
\affiliation[f]{University of Geneva, section de Physique, DPNC, Geneva, Switzerland}

\affiliation[g]{LPNHE Paris, Sorbonne Universite, Universite Paris Diderot, CNRS/IN2P3, Paris, France}

\affiliation[h]{Institut de F\'isica d’Altes Energies (IFAE) - The Barcelona Institute of Science and Technology (BIST), Campus UAB, 08193 Bellaterra (Barcelona), Spain}
\affiliation[i]{University of Pennsylvania, Department of Physics and Astronomy, Philadelphia, PA 19104, USA}
\affiliation[j]{Institute for Nuclear Research of RAS, Moscow, Russia}
\affiliation[k]{ State University of New York at Stony Brook, Department of Physics and Astronomy, Stony Brook, New York, U.S.A.}
\affiliation[l]{University of Tokyo, Department of Physics, Tokyo, Japan}
\affiliation[m]{High Energy Accelerator Research Organization (KEK), Tsukuba, Japan}

\affiliation[n]{ETH Zurich, Institute for Particle Physics and Astrophysics, CH-8093 Zurich, Switzerland}
\affiliation[o]{Louisiana State University, Department of Physics and Astronomy, Baton Rouge, Louisiana, USA}
\affiliation[p]{Moscow Institute of Physics and Technology (MIPT), Moscow Region, Russia}
\affiliation[q]{National Research Nuclear University MEPhI, Moscow, Russia}
\affiliation[r]{University of Pittsburgh, Pittsburgh, PA, 15260, USA}

\emailAdd{skrobovana@lebedev.ru}

\abstract{

The SuperFGD detector will be a novel and important upgrade to the ND280 near detector for both the T2K and Hyper-Kamiokande projects. The main goal of the ND280 upgrade is to reduce systematic uncertainties associated with neutrino flux and cross-section modeling for future studies of neutrino oscillations using the T2K and Hyper-Kamiokande experiments. The upgraded ND280 detector will be able to perform a full exclusive reconstruction of the final state from neutrino-nucleus interactions, including measurements of low momentum protons, pions and for the first time, event-by event measurements of neutron kinematics. Precisely understanding the time resolution is critical for the neutron energy measurements and hence an important factor in reducing the systematic uncertainties. In this paper we present the results of time resolution measurements made with the SuperFGD prototype that consists of 9216 plastic scintillator cubes (cube size is 1~cm$^3$) readout with 1728 wavelength-shifting (WLS) fibers along the three orthogonal directions. We used data from a muon beam exposure at CERN. A time resolution of 0.97~ns was obtained for one readout channel after implementing the time calibration with a correction for time-walk effects. The time resolution improves with increasing energy deposited in a scintillator cube, 
improving to 0.87~ns for large pulses. 
Averaging two readout channels for one scintillator cube further improves the time resolution to 0.68~ns implying that signals in different channels are not synchronous. In addition the contribution from the time sampling interval of 2.5~ns is averaged as well. Most importantly, averaging time values from N channels improves the time resolution by $\sim 1/\sqrt{N}$. For example, averaging the time from 2 scintillator cubes with 2 fibers each improves the time resolution to 0.47~ns which is much better than the intrinsic electronics time resolution of 0.72~ns in one channel due to the 2.5~ns sampling window. This indicates that a very good time resolution should be achievable for neutrons since neutron recoils typically interact with several scintillator cubes and in addition produce larger signal amplitudes than muons. Measurements performed with a laser and a wide-bandwidth oscilloscope in which the contribution from the electronics time sampling window was removed demonstrated that the time resolution obtained with the muon beam is not far from the theoretical limit. The intrinsic time resolution of a scintillator cube and one WLS fiber is about 0.67~ns for signals of 56~photo electrons which is typical for minimum ionizing particles.

}


\begin{document}
\maketitle
\flushbottom

\section{Introduction}

The SuperFGD detector will be a novel and important upgrade to the ND280 near detector for both the T2K and Hyper-Kamiokande projects. The main goal of the ND280 upgrade~\cite{UpgradeTDR} is to reduce systematic uncertainties associated with neutrino flux and cross-section modeling for future studies of neutrino oscillations using the T2K and Hyper-Kamiokande experiments. The upgraded ND280 detector will be able to perform a full exclusive reconstruction of the final state from neutrino-nucleus interactions, including measurements of low momentum protons, pions and for the first time, event-by event measurements of neutron kinematics.
The physics sensitivity that can be expected from the upgraded detector is described in~\cite{PhysRevD.105.032010}.
Precise measurements of the outgoing neutron could improve antineutrino energy reconstruction~\cite{NeutronEnergyRec}. This method relies on accurate neutron kinematics reconstruction. Neutron initial energy could be calculated using Time-of-Flight measurements between an antineutrino interaction vertex and hits caused by a secondary proton or other particles produced by the neutron. Good time resolution is crucial for such types of measurements.
The Super Fine-Grained Detector (SuperFGD) with its quasi-3D readout scheme is a key component of the T2K ND280 near detector upgrade~\cite{Blondel:2017orl}.
In this paper we present the results of time resolution measurements made with the  SuperFGD prototype exposed to a muon beam at CERN. The SuperFGD prototype tests using charged particle beams at CERN are described elsewhere~\cite{SFGD_CERN_test}. Here we use the same data (2~GeV muons) to estimate the ultimate SuperFGD time resolution. The results from independent tests with a laser light source and a wide-bandwidth oscilloscope are also presented.

\begin{figure}[h]
\centering
\includegraphics[width=0.52\textwidth]{detector.pdf}
\includegraphics[width=0.4\textwidth]{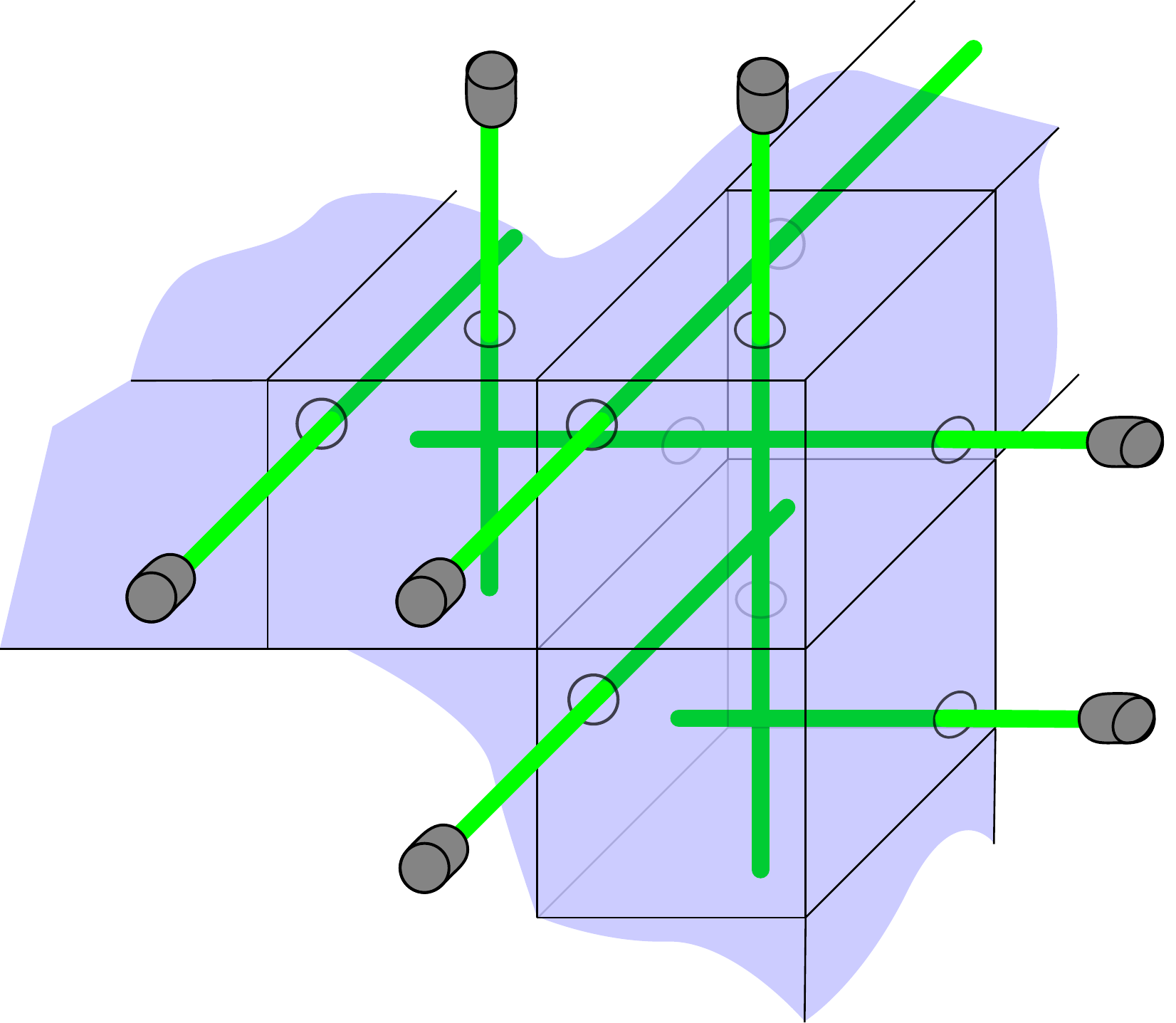}

\caption{Illustration of the detector structure. Left: Distribution of the three types of MPPCs (Type I -- S13360-1325CS, Type II -- S13081-050CS, and Type III -- S12571-025C) around the six faces of the SuperFGD prototype. In the CERN test the beam was parallel to the Z axis. 
Right: a schematic view of the detector corner with SiPM readout and scintillator cubes with the WLS fibers going through the cubes.
}

\label{fig:sfgd}
\end{figure}

The SuperFGD prototype dimensions are $24 \times 8 \ \times 48$ cm$^3$, (see figure~\ref{fig:sfgd}). It consists of 9216 plastic scintillator cubes (the cube size is 1~cm$^3$). Three wavelength-shifting fibers along three orthogonal directions penetrate each cube. One end of each fiber is read out with a Silicon Photo-Multiplier (SiPM).
Three types of SiPMs are used in the Prototype: Hamamatsu MPPC S13360-1325CS, S13081-050CS, and  S12571-025C (see figure~\ref{fig:sfgd} for their positions in the prototype).  
Their properties are described in table~1~\cite{SFGD_CERN_test}. 
Type~I MPPC (S13360-1325CS) was selected for the final SFGD design. Two other types of the SiPMs were installed in order to save on instrumentation costs (they were available as spares).
There are 
1728 fibers and readout channels. The design of the SuperFGD prototype is very similar to the design of the full SuperFGD. However, the SuperFGD will be much larger. It will consist of 192 (width) × 56 (height) × 182 (length) scintillator cubes, where each cube is 1~cm$^3$.
The SuperFGD prototype electronics is the same as in the final SuperFGD design. It is based on the CITIROC (Cherenkov Imaging Telescope Integrated Read Out Chip) front end ASIC (Application-Specific Integrated Circuit), which is designed for SiPM read out~\cite{Electronics18, Noah:20165S}. Time measurements are performed using a fast shaper with a fixed threshold discriminator. Typical thresholds correspond to about 3 photo electrons (p.e.).
The Time-over-Threshold (ToT) signal is used for amplitude measurements since it is recorded continuously without any dead time (unlike High Gain and Low Gain channels)~\cite{SFGD_CERN_test}.

Energy released by a charged particle in a scintillator is transferred first to the first dye (p-Terphenil). The light emitted by the first dye (maximal intensity at $\sim$340~nm) is absorbed by the second dye (POPOP) and then emitted at a longer wavelength (maximal intensity at $\sim$410~nm) close to the maximum of the WLS fiber absorption spectrum. The WLS fiber shifts the spectrum close to the maximum of the SiPM spectral sensitivity (maximal intensity at $\sim$450~nm). Fluctuations in all these processes contribute to the
fluctuations in the time measurements.
There are also fluctuations in the SiPM response, but they are much smaller. Contributions from all these stages can be considered as the intrinsic time resolution of the whole scintillator counter. Due to fluctuation in the de-excitation of the dyes and  the WLS fiber the photoelectrons arrive later than the fastest possible arrival time.
The time jitter on the arrival of the first photoelectron decreases with an increasing number of photons produced. Therefore, the time resolution improves with the number of produced photoelectrons. The readout electronics also contribute to the time resolution.
The time digitization step in the SFGD electronics is 2.5~ns. 
Therefore, the time resolution of one channel cannot be better than 2.5~ns/$\sqrt{12}$ (the standard deviation of the uniform distribution within the 2.5~ns bin). With a fixed threshold the measured time depends on the signal amplitude. Therefore, a corresponding time-walk correction is required. 
In this paper we investigate contributions of the different processes to the time resolution by exciting the first and second dyes as well as the WLS fiber with a laser light and compare the results with the time resolution obtained for muons.

\section{Time resolution of individual channels}
\label{sec:resolution_one_channel}

To measure the ultimate time resolution, muon tracks going strictly parallel to the Z-axis were used. 
Although the cube surface is covered with a reflecting layer, this layer is not completely opaque. Some hits associated with the muon track can therefore be caused by optical crosstalk, i.e. light passing between neighboring cubes. Crosstalk is also possible through the fibers, though this effect is much smaller~\cite{xtalk}.

\begin{figure}[h]
\centering
\includegraphics[width=0.48\textwidth]{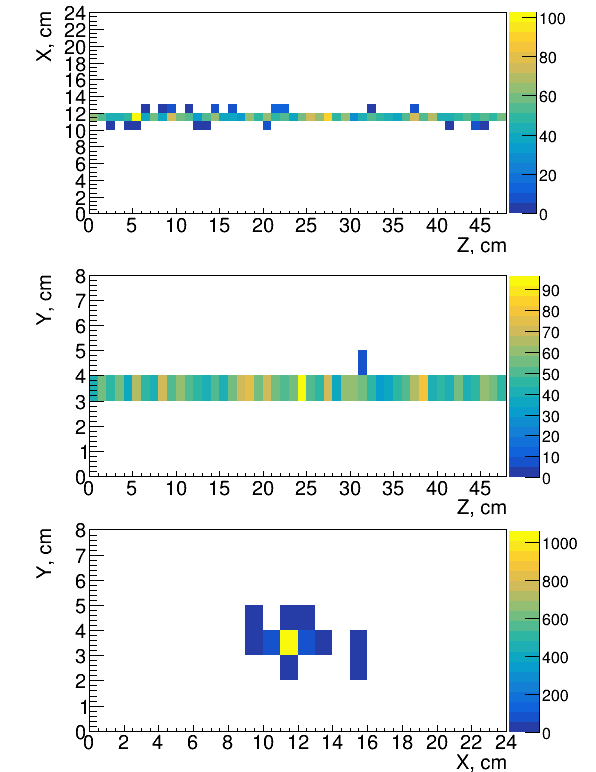}
\includegraphics[width=0.48\textwidth]{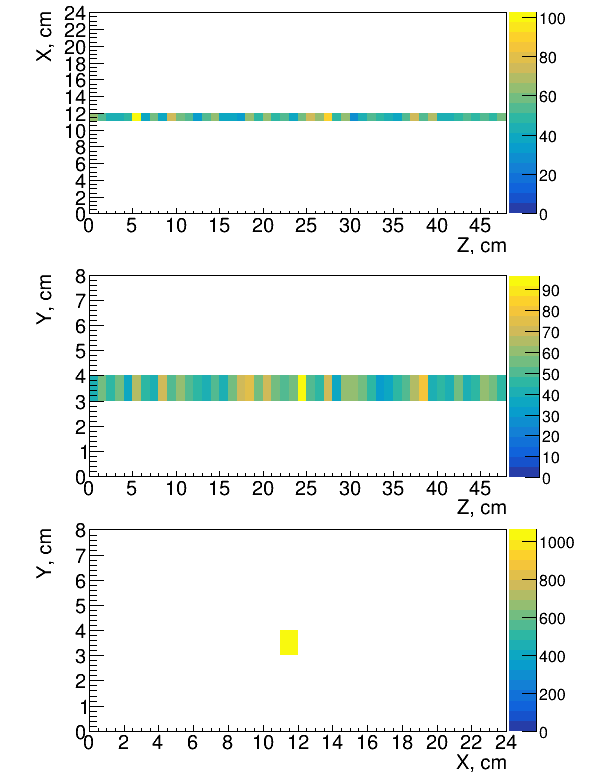}

\caption{Illustration of the track reconstruction procedure. Left: the distribution of hits in an event in 3 projections.
Right: the distribution of cubes selected on the track in 3   projections after the removal of crosstalk. For all distributions color corresponds to the light yield in p.e. 
X coordinates of hits in the XZ projection should match the X coordinate of the hit with the highest amplitude in the XY projection (``yellow'' in the picture). There are similar requirements for Y coordinates in the YZ projection. The cube is considered to be reconstructed only if it has signals from both X and Y fibers.} 

\label{fig:event_selection}
\end{figure}

In this paper a hit denotes a signal from a single WLS fiber.
In this study we only consider the X and Y fibers for time measurements. We don't consider Z fibers for time measurements because in this case the energy released in several cubes contributes to one single hit. In order to achieve the best time resolution, hits caused by  crosstalk are not used in the analysis.
In the XZ plane we select hits with the X coordinate that matches the X coordinate of the hit with the highest amplitude in the XY plane (exposed to the beam) and similarly for the hits in the YZ plane we select hits with the Y coordinate that matches the Y coordinate of the hit with the highest amplitude in the XY plane.
For this analysis we select more than 40 hits associated with the track both in the XZ plane and in the YZ plane.
Then the 3D structure of the track is reconstructed (see figure~\ref{fig:event_selection}). 
Since crosstalk signals cannot produce the highest amplitude hit in the XY plane,
optical crosstalk to cubes outside the selected track position is effectively removed 
with the exception of the tracks which cross the border between two cube columns along the Z fibers. However, the number of such tracks is negligible.
The crosstalk between the cubes within the track is not important because the light from the crosstalk is synchronous or slower than direct light from the muon, which sets the timing reference. The crosstalk signals are much smaller than the signals caused by muons.

\begin{figure}[h]
\centering
\includegraphics[width=0.6\textwidth]{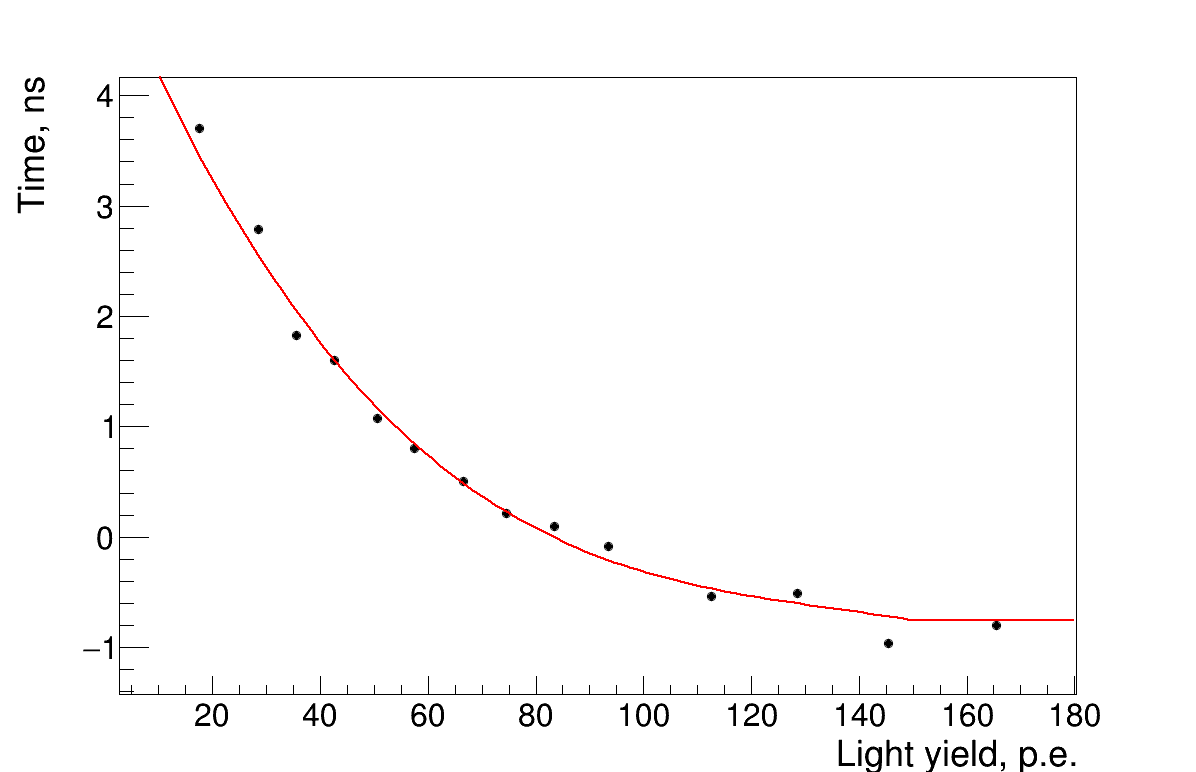}
\caption{Example of the measured difference between the hit time and the reference time as a function of the light yield for a Y fiber of one cube. The average time of all hits with more than 20~p.e. is used as the reference time. Light yield data are fitted with a third-order polynomial function. Fit range is 20 -- 150~p.e.  
For higher amplitudes a constant extrapolation is used. Higher amplitudes are not used in the fit because of low statistics. Small amplitudes are not used in the time resolution studies. 
} 
\label{fig:timewalk_cube}
\end{figure}

The time difference between the signal arrival to the discriminator and the moment when the signal reaches a given detection threshold depends on the number of photons in the light pulse.
Signals with larger amplitudes have a faster rise time. Due to the time-walk effect hits with lower amplitudes have later values of recorded times. 
As an example, the dependence of the hit time on amplitude for a Y fiber (i.e. a fiber parallel to the Y-axis) of a single cube is presented in figure~\ref{fig:timewalk_cube}. 

The absolute hit times were measured with respect to the beam spill signal. However, a single beam spill would typically contain many muon events spread over the time of the spill. Therefore, the start signal of the spill alone could not provide a good time reference for measuring the times of the hits relative to each muon. Instead, we measured the time differences between each individual hit and the mean event time of the muon track determined
by averaging all track hits above 20~p.e.
Hence we determine the event time as the average time of all hits with more than 20~p.e. in the reconstructed muon track.

The time-walk correction was measured for every single cube relative to the reconstructed average time of the muon tracks. Third-order polynomial functions  were used to fit the data. Errors of all points are assumed equal in the fit to take into account possible systematic uncertainties. These functions were used to correct time for every cube independently for the Y and X fibers.
The amplitude-dependent terms of these functions address the time-walk correction. The constant terms of these functions correspond to time shifts in individual MPPC channels, time delays due to different distances from the cubes to MPPCs, and  muon travel times.
The here described time calibration procedure automatically takes into account the light propagation time along the fiber and the muon travel time along the detector.

\begin{figure}[h]
\centering
\includegraphics[width=0.8\textwidth]{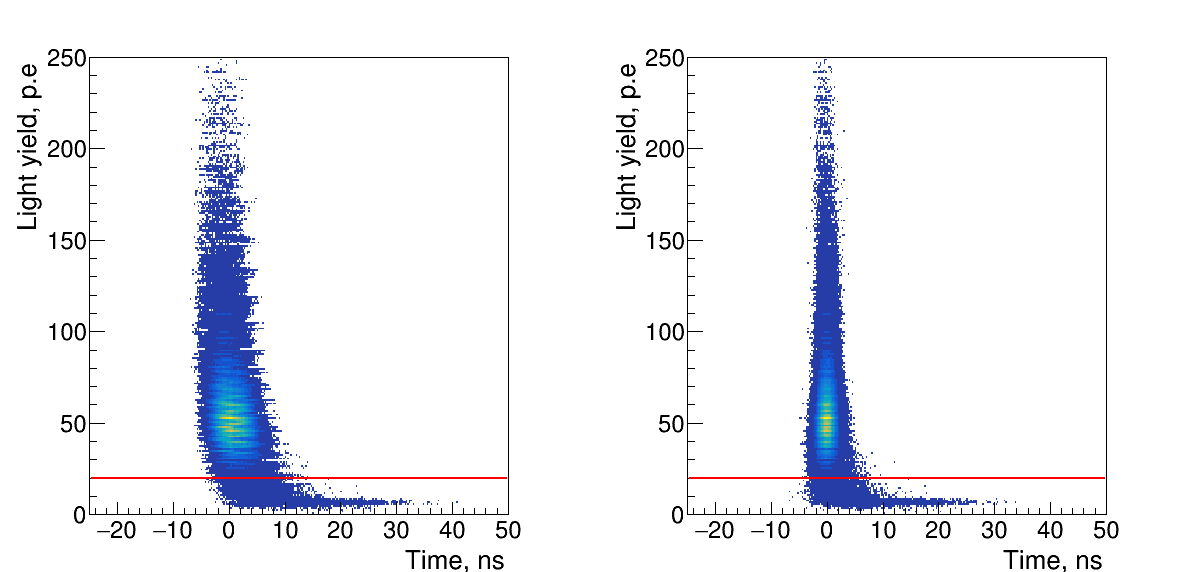}
\caption{Hit amplitude vs hit time (hit time minus reference time) distribution before (left) and after (right) the time-walk correction for X fibers. 
The average time of all hits with more than 20~p.e. is used as the reference time. Hits with large time values are dominated by the remaining crosstalk. The time walk correction was determined using signals with more than 20~p.e. Therefore, the correction for signals with very small amplitudes is not complete.
In time resolution studies only hits with amplitudes higher than 20~p.e. are considered (values above the red line).
}
\label{fig:timewalk_not_corr}
\end{figure}

The results of  the implementation of the time-walk correction are illustrated in figure~\ref{fig:timewalk_not_corr}.
The time-walk corrections are validated by applying them to a different data sample with the same event selection and beam conditions, and observing the improvement in overall time resolution.
Only the central part of the SuperFGD Prototype is used where a lot of tracks parallel to Z can be reconstructed, namely, the tracks passing through cubes with X=10-12 and Y=3-4 are included in the analysis. 
All time resolution estimates are made for hits with an amplitude greater than 20~p.e.
This requirement removes less than 5\% of hits.

\begin{figure}[h]
\centering
\includegraphics[width=0.32\textwidth]{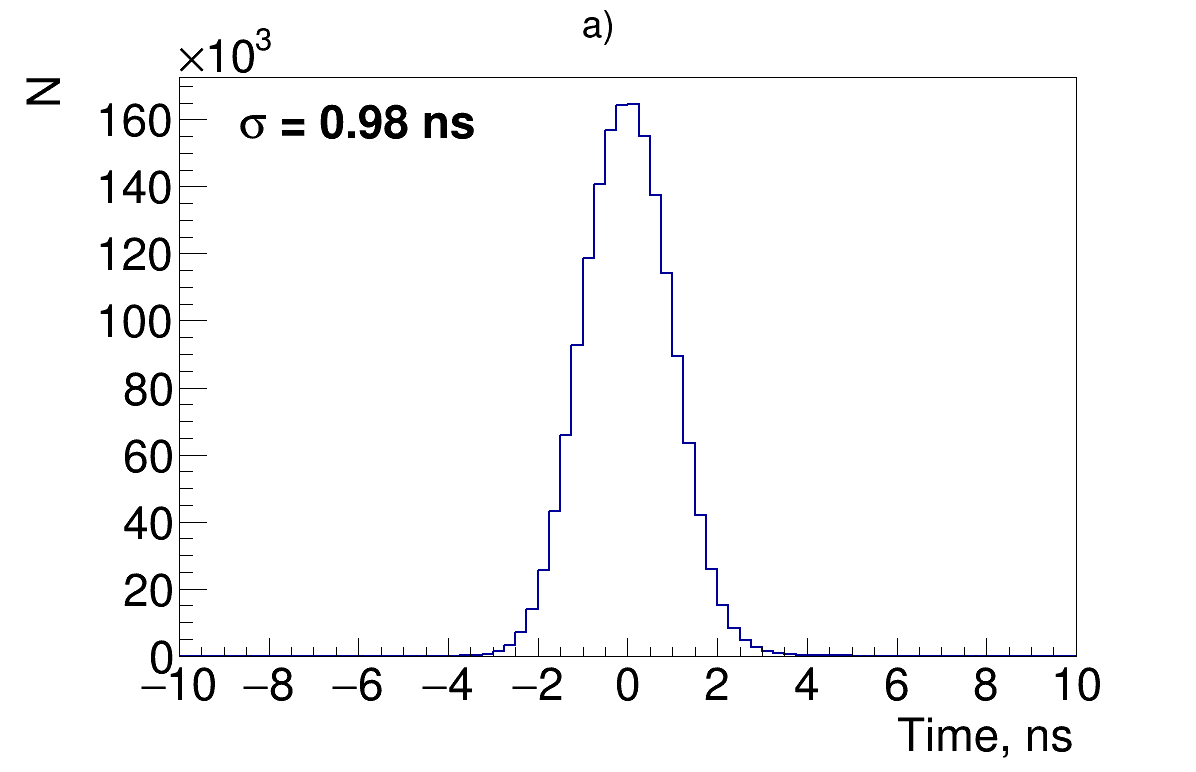}
\includegraphics[width=0.32\textwidth]{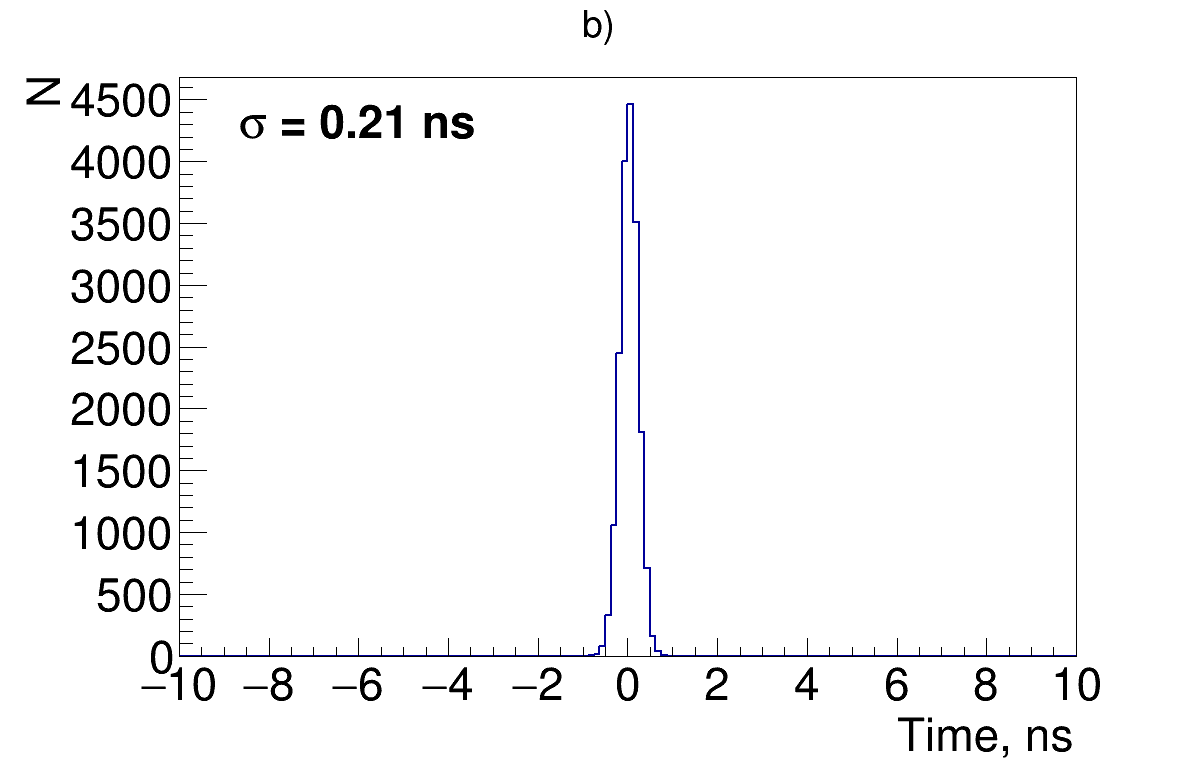}
\includegraphics[width=0.32\textwidth]{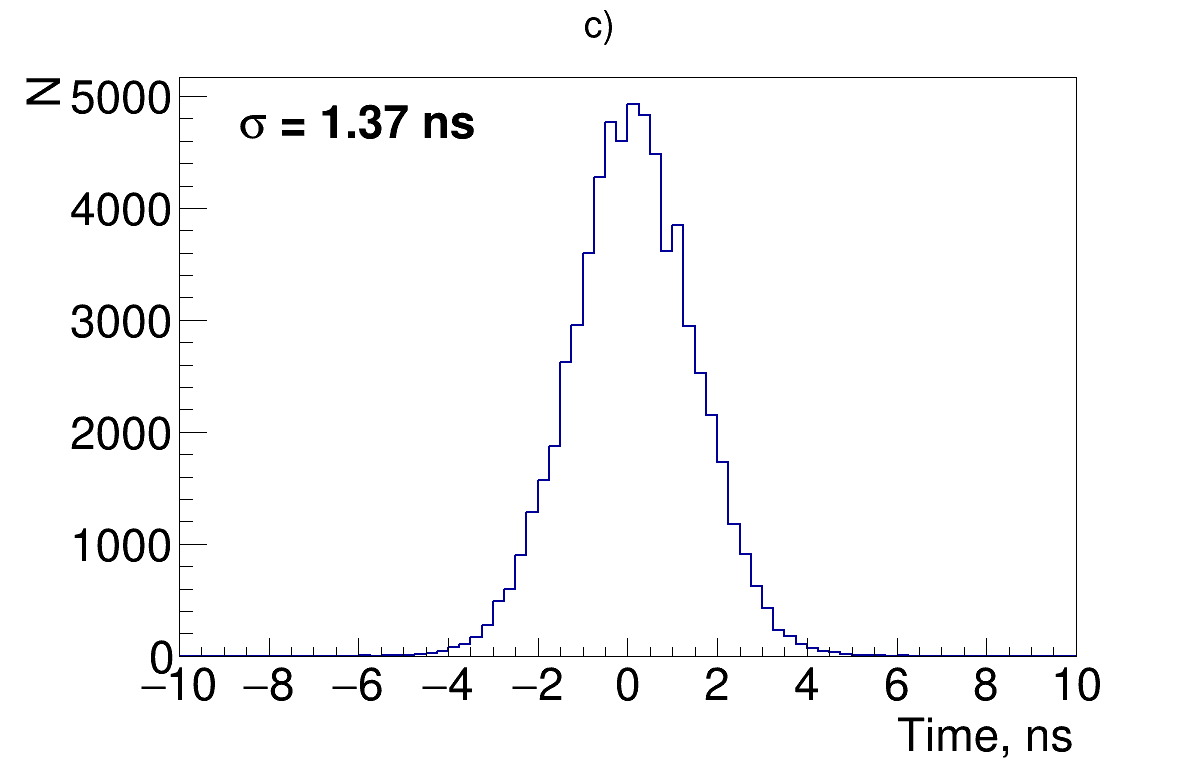}

\caption{a): The time (hit time minus reference time) distribution of hits in events. The event average time is used as a reference time. b): The distribution of the difference between the event average times calculated using the odd and even layers. c): The distribution of the difference between two random hit times in the same event. } 
\label{fig:time_diff}
\end{figure}

The time resolution for all individual channels is measured using the time distribution of hits in the track (from both X and Y fibers) with respect to the average time of each event (see figure~\ref{fig:time_diff}~a). The standard deviation of this distribution is $\sigma = 0.98~$ns. Two parts contribute to the standard deviation of this distribution. The first one is the intrinsic time resolution of one channel. The second one is the resolution of the time reference, i.e. of the event average time. It is determined in the following way.
One can calculate the average time of an event using only half of the hits in the muon track. The accuracy of this estimate is $\sqrt{2}$ times worse than the resolution in the average event time based on all hits in the muon track provided that the time measurements in different channels have the same resolution and that they are randomly distributed. We will use these assumptions in all estimates of the time resolutions. The validity of these assumptions is confirmed by the consistency of the obtained results. The distribution of the difference between event average times calculated using odd and even layers is presented in figure~\ref{fig:time_diff}~b). The standard deviation of this distribution is $\sqrt{2}$ times bigger than the standard deviation of the  average time calculated using half of the hits in the event. Therefore, the event average time ($t_0$) resolution is 2 times better: $\sigma_{t_0} = \frac{1}{2}\sigma_{odd-even} = 0.1~ \mathrm{ns}$. Hence, the intrinsic time resolution of one channel ($t_1$) is equal to:
$$
\sigma_{t_1} = \sqrt{\sigma_{t_0-t_1}^2 - \sigma_{t_0}^2} = 0.97~\mathrm{ns},
$$
where $t_0$ is the reference time.
The statistical accuracy of this estimate is better than 0.01~ns.

The intrinsic time resolution of one channel can be also estimated in a different way using the  distribution of the time difference between 2 random hits in the same event (figure~\ref{fig:time_diff}~c).  The standard deviation of this distribution is $\sqrt{2}$ larger than the time resolution of a single hit.
According to this estimate, the single channel time resolution is 0.97~ns in good agreement with the result obtained with the previous method. This coincidence of the two estimates indicates that  
the systematic errors 
in  the time resolution determination
are small. The statistical errors are negligible.

The obtained time resolution is 18\% better than the time resolution (1.14 $\pm$ 0.06~ns) obtained in the same beam tests in the previous analysis~\cite{SFGD_CERN_test}. 
This is most likely due to an improved calibration procedure, removal of crosstalk, and implementation of the 20~p.e. threshold for the hit amplitude.

\section{Time resolution dependence on amplitude}

\begin{figure}[h]
\centering
\includegraphics[width=0.7\textwidth]{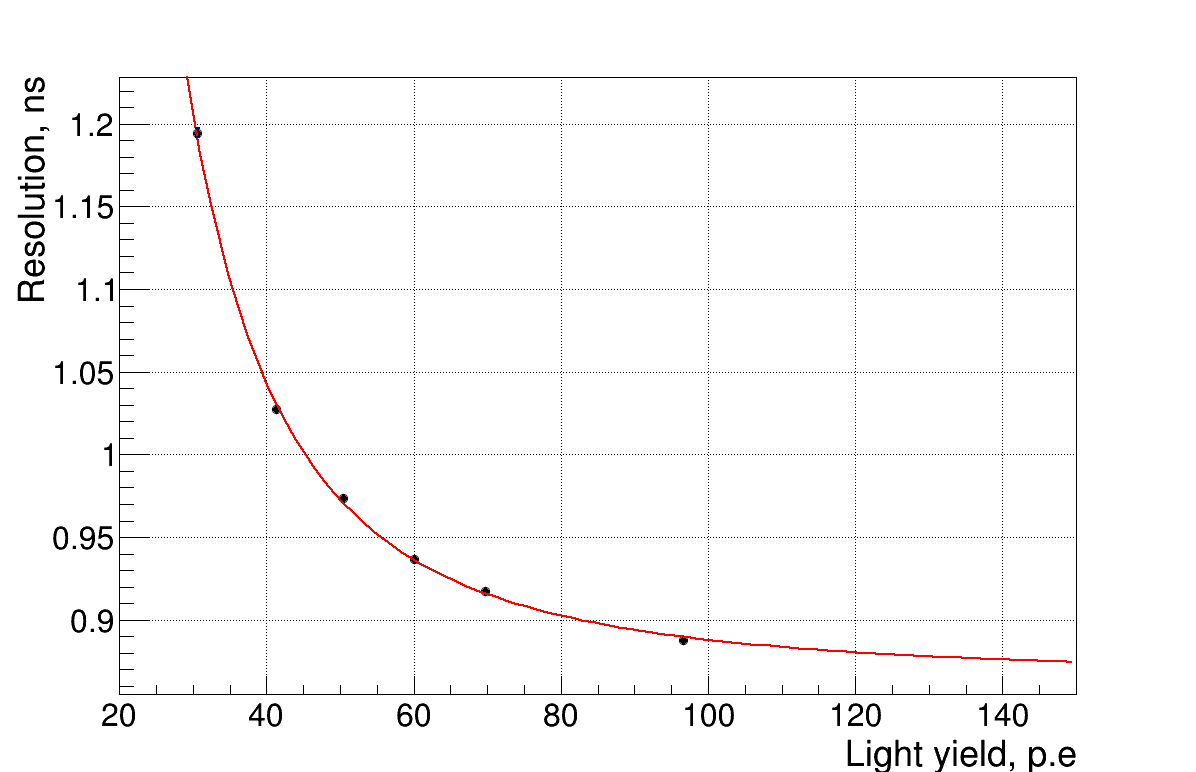}
\caption{Time resolution dependence on the light yield for one hit. Only statistical errors are shown. 
} 
\label{fig:energy_time_resolution}
\end{figure}

The time resolution depends on the signal amplitude. Therefore, the time resolution of one channel was measured in different amplitude intervals independently. The same algorithm was used as described in section~\ref{sec:resolution_one_channel}.
Results are presented in figure~\ref{fig:energy_time_resolution} for Type-I MPPCs  which were selected for the final SFGD design. They have a large number of pixels (2668) and a small crosstalk ($\sim 1\%$). Therefore, the number of fired pixels and photoelectrons, in case of small signals, are practically identical.
The SiPM contribution to the time resolution is negligible (see the discussion in section~\ref{sec:laser}).
The time resolution improves with amplitude up to 0.89 ns for amplitudes between 80 and 250~p.e..
The average signal value for the full interval (20–250~p.e.) used in this analysis is 56~p.e.
 The following function was used to fit the data: 
 $$\sigma(A) = p_0/A^{p_1}+p_2,$$ 
 where $\sigma$ -- is time resolution in ns, A -- is amplitude in p.e., $p_0, p_1$, and $p_2$ are free parameters. The value of $p_2$ gives the high amplitude limit on the time resolution of $p_2 = 0.87\pm0.01$~ns. 
This value is somewhat larger than the expected limit from the digitization step of 2.5~ns/$\sqrt{12}$ = 0.72~ns.
This suggests that either the amplitude range is not sufficient to determine reliably the 
 high amplitude limit on the time resolution or the amplitude dependence is not well described by the assumed functional form.

\section{Averaging of different channels}

It was demonstrated in section~\ref{sec:resolution_one_channel} that the event average time resolution is much better than the time resolution of a single hit. 
This means that averaging of different channels improves the time resolution. This is by far not a trivial fact. There would be no such improvements if the causes of the time fluctuations were correlated between hits
or if they were much smaller than the effect of the digitization step. 
If all effects causing time fluctuations were much smaller than the effect of the time digitization, no improvement beyond the time digitization limit would be possible.
In such a case the time resolution after averaging several channels would be limited by the uncertainty due to the time digitization step of 2.5~ns/$\sqrt{12}$ = 0.72~ns.

SuperFGD electronics allow a very good synchronization of different channels. The time delay difference between front-end boards of  0.07~ns was obtained in a test setup~\cite{FEB_timing}. 
However, time measurements in different channels are smeared out by several effects, and hence they are  distributed in several (typically 2 or 3) 2.5~ns time bins. 

The following effects smear the time measurements.  The intrinsic time resolution of the scintillator cube and one WLS fiber is about 0.7~ns for signal amplitudes typical for muons (this will be discussed in the next section). 
Different ToT thresholds and/or amplifications in different channels will also influence the time measurements. The time-walk correction varies from 0~ns to 2~ns for signals typical for muons (see figure~\ref{fig:timewalk_cube}). 
Finally, the time delays caused by a muon movement through the detector can also shift the measured time to different time bins. The last two effects are corrected later in the analysis. 
However, they cause shifts in the raw time measurements.
All four of these effects are not correlated.
The spread in the time measurements caused by these effects is comparable with the width of the time bin of the readout electronics. Therefore, the time resolution is improved relative to the intrinsic time resolution of the readout electronics after averaging several channels.

\begin{figure}[h]
\centering
\includegraphics[width=0.48\textwidth]{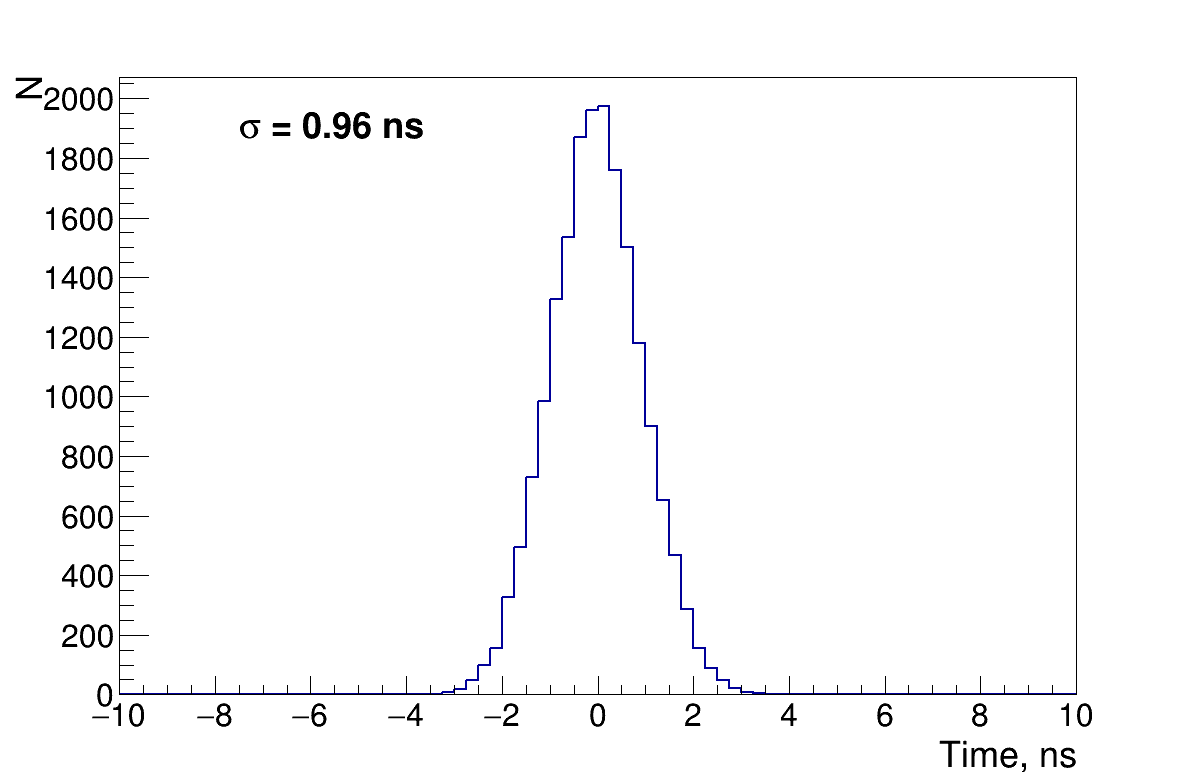}
\includegraphics[width=0.48\textwidth]{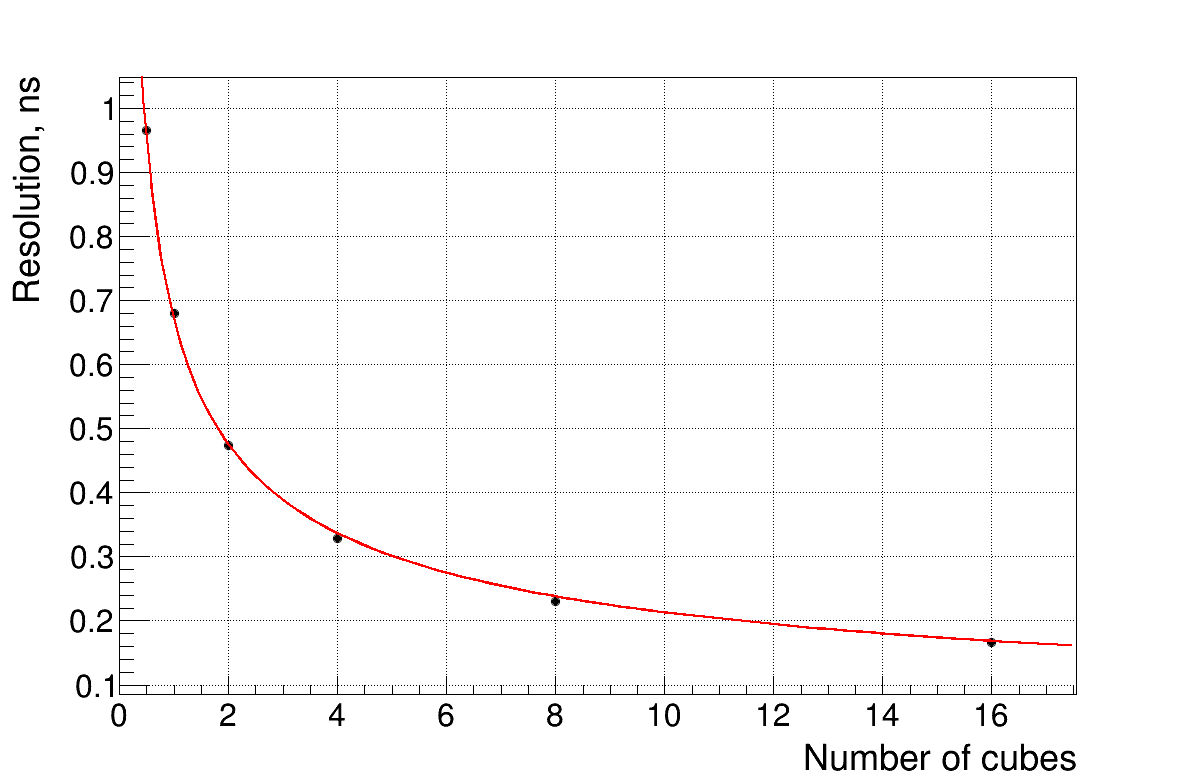}

\caption{Left panel:  Distribution of the difference between two scintillator cube times in the same event that has $\sigma = 0.96$~ns which is $\sqrt{2}$ times larger than the standard deviation of the one cube time resolution. The cube time is the average of the corresponding 
times measured with X and Y fibers. Right panel: The time resolution dependence on the number of cubes (N) used in the time averaging. In addition, the  resolution of a single hit is demonstrated (the first point at N=0.5).
The curve clearly demonstrates the $1/\sqrt{N}$ dependence. For the range of cubes fitted (N$\leq$16) the expected constant term is consistent with zero and therefore not included in the fit.
} 
\label{fig:unify}
\end{figure}

Due to these effects
averaging a different number of N channels in the muon track leads to the improvement in the time resolution $\sim 1/\sqrt{N}$. 
Figure~\ref{fig:unify} (left) shows the distribution of the time difference between 2 random cubes in the same event where the cube time is the average time measured by X and Y fibers 
: $t_{cube} = (t_X + t_Y)/2$. The standard deviation of this distribution is $\sqrt{2}$ larger than the time resolution of a single cube. 
The obtained time resolution of a single cube is 0.96~ns/$\sqrt{2}$ = 0.68 ns in good agreement with the estimate of 0.97~ns$/\sqrt{2} = 0.69$~ns based on averaging of two independent channels.
In the same way, the time resolutions for 2, 4, 8, and 16 cubes averaged were obtained. 
The results are presented in figure~\ref{fig:unify} (right). The obtained improvement of the time resolution on the number of cubes used in the time averaging is in a reasonable agreement with the $1/\sqrt{N}$ law.

These improvements from averaging multiple channels mean that a very good time resolution should be achievable for neutrons since neutron recoils typically hit several cubes and in addition produce larger amplitudes than muons.
In addition for neutrons, time signals from three fibers per cube  can be used for averaging while in the present study only X and Y fibers were used.

\section{Time resolution studies with a laser}
\label{sec:laser}

\begin{figure}[h]
\centering
\includegraphics[width=0.52\textwidth]{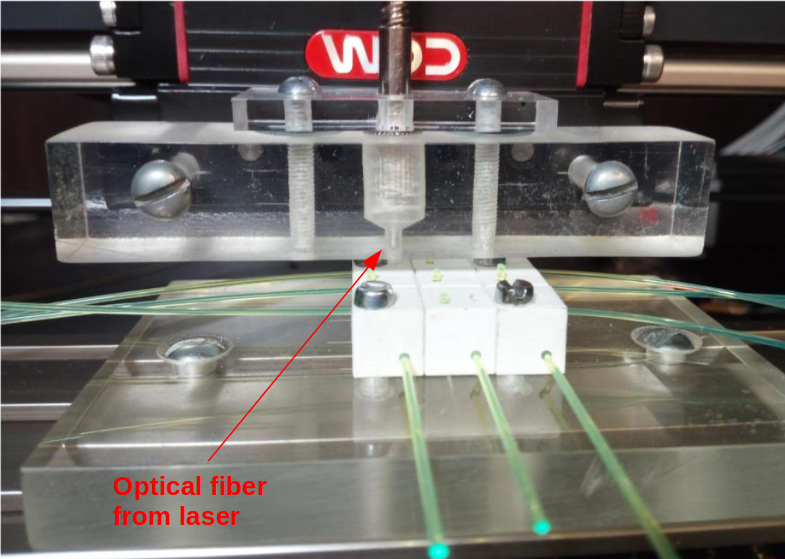}
\caption{Experimental setup with 9 scintillator cubes, 6 readout fibers and an optical fiber from a laser. 
} 
\label{fig:laser}

\end{figure}

An independent study using a 266~nm wavelength laser source (UV picosecond laser VisUV-266-355-532 by PicoQuant GmbH) was performed to measure the time resolution with a high-frequency digitizer of the LeCroy WavePro 254HDR Oscilloscope to avoid the contribution from the 2.5~ns digitization step of the SuperFGD electronics. 
Nine scintillator cubes were arranged in a layer with three readout WLS fibers along X (left-right) and three fibers along Y in the layer X-Y plane 
(see figure~\ref{fig:laser}). Short WLS fibers were inserted in the vertical cube holes but were not read out. The readout WLS fibers had 3 lengths: 15~cm, 25~cm, and 134~cm. 

For time resolution studies the scintillator cubes were illuminated with the 266~nm laser via an optical fiber through the reflective cube surface. 
This wavelength  is close to the maximal absorption point of the first scintillator dye p-Terphenil.
The wave-forms of signals recorded by the oscilloscope were analysed using a constant fraction threshold algorithm. A few measured points on the signal front close to the threshold value were fit with a straight line to determine the time at threshold. Signals with a large dispersion of the baseline measurements and/or large $\chi^2$ of the straight line fit  were discarded to avoid the influence of noise. The measured difference  between the laser pulse time and the time at a selected threshold was fit with a Gaussian function in order to exclude the contribution from the distribution tails. The standard deviations obtained from the fit were about 5--10\% better than the standard deviations calculated directly from the measured distributions.   The best time resolutions were obtained with the threshold amplitude of about 10\%  of the first maximum of the wave-form. The thresholds of 6\%, 8\%, 20\%, 30\%, and 50\% gave worse time resolutions for signals from a cube illuminated with 40~p.e.

\begin{figure}[h]
\centering
\includegraphics[width=0.62\textwidth]{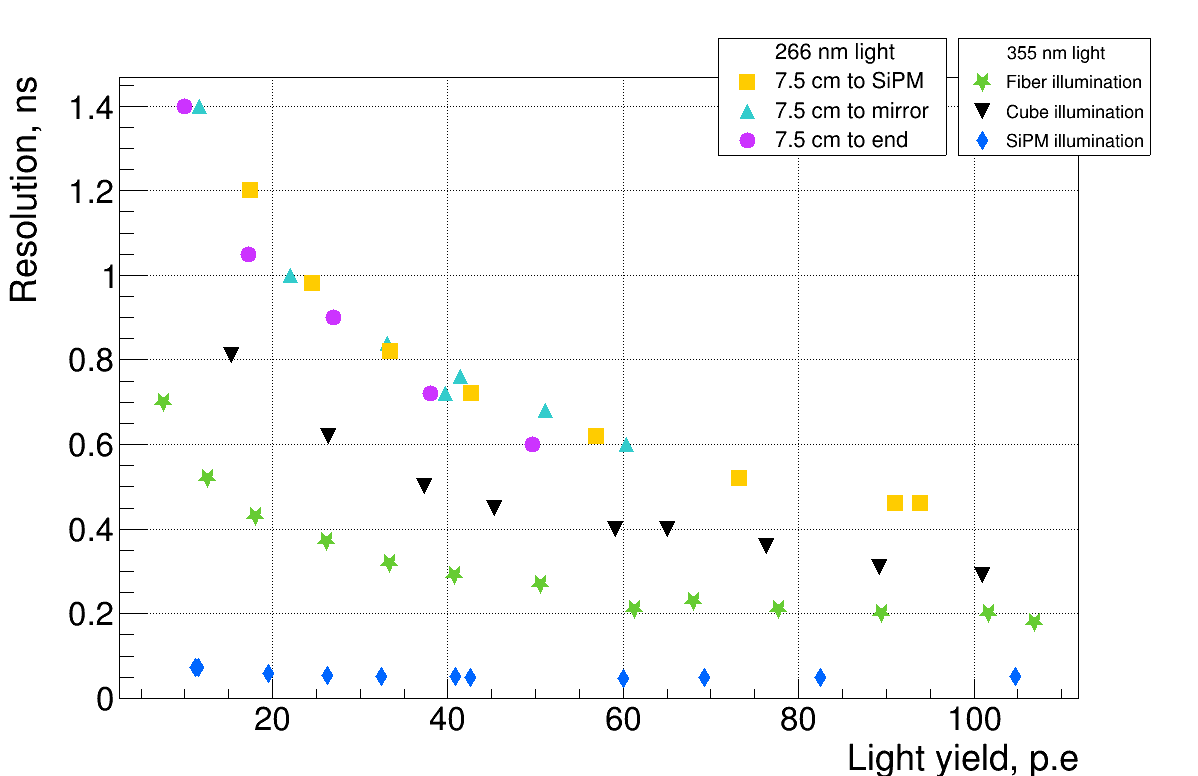}
\caption{Time resolution dependence on the light yield obtained for different conditions. For the 266~nm laser light the following cases are shown. Square markers: the cube position was 7.5~cm to the SiPM, triangle up markers: the cube position was 7.5~cm to the fiber end with the silver paint, circle markers: the cube position was 7.5~cm to the fiber end without the silver paint. For the 355~nm laser light the three cases are shown: cube illumination (triangle down markers), fiber illumination (star markers), and direct SiPM illumination (diamond markers). Fiber and cube measurements with the 355~nm light were performed at the 7.5~cm distance from the SiPM. For all measurements the statistical errors are smaller than marker sizes. Systematic errors were evaluated by reproducing measurements with the same conditions. They were found to be at the level of 5--10\%.
} 
\label{fig:laser_resolution}
\end{figure}

The output of the light emitting fiber could be moved over the surface of the cubes using two stepper motors controlled by Arduino microcontroller.
The average value of the response signal varied over the scintillator cube surface probably due to different thickness and/or quality of the reflective layer.
The time resolution was anti-correlated with changes of the light yield.
However, these variations did not influence the results since the time resolution dependence on the signal charge was measured with a fixed position of the laser light-emitting fiber at the center of the central cube.
Non-illuminated cubes were covered with black paper to reduce the influence of reflections.
Fibers were read out with  Hamamatsu MPPCs S13360-1350CS 
and  preamplifiers described in~\cite{DANSS_digitization}.
The SiPMs used in the SFGD Protptype were not available to us during the measurements with the laser. However, the results do not depend on the SiPM type since fluctuations in the time measurements caused by SiPMs are negligible in comparison with other effects (see figure~\ref{fig:laser_resolution}).
The gain calibration and crosstalk determination were done using peaks corresponding to 1 and 2 fired pixels. 
The light yield was calculated from the observed number of fired MPPC pixels using corrections for the crosstalk of about 3\% and for the  saturation due to the limited number of pixels in the MPPC (667).
The contribution of the MPPC and electronics to the time resolution was determined by directly illuminating MPPC with the 355~nm laser light. The obtained time resolution varied from 0.07~ns for 10~p.e. signals to 0.05~ns for signals larger than 40~p.e.
Therefore, in this study the time resolution is dominated by fluctuations in the light emission from the scintillator and WLS fibers.

The obtained time resolution dependence on the light yield is presented in figure~\ref{fig:laser_resolution}. 
Measurements were performed illuminating the scintillator cube center with the light collection using a 134~cm long WLS fiber. In one measurement the scintillator cube was about 7.5~cm from the MPPC while in two other measurements it was about 7.5~cm from the fiber far end that is about 126.5~cm from the MPPC. The fiber far end was mirrored with a silver paint. In the last measurement a few millimeters of this mirrored fiber end were cut away. 
The obtained dependencies of the time resolution on the light yield were similar in all 3 measurements.
For the same number of detected p.e. we have not observed a sizable difference in the time resolution for measurements near the photo-detector and at the fiber far end as well as for the mirrored and cut fiber ends.

The measurements performed close to the mirrored fiber end had very similar  conditions with  the measurements performed using  the small 5x5x5~cm$^3$ prototype  with a 5 GHz digitizer at the CERN beam tests\cite{SFGD5x5}. The obtained time resolution of ~0.8~ns for  40~p.e. signals was slightly better than 0.95~ns obtained at the CERN beam tests~\cite{SFGD5x5}. There were several differences between these two measurements. 
The beam measurements were not corrected for the time fluctuations of the trigger counters. However, this contribution was small. The time resolution determined from the difference of times from two fibers from the same cube was 0.92~ns. This method of time resolution determination did not rely on the trigger time measurements. Also the distributions of the number of p.e. in the two tests were different in spite of the same average values.
In the beam tests Landau fluctuations resulted in a wide distribution with a long tail at large number of p.e. Hence the bulk of events had smaller number of p.e. than the average number.

For a better understanding of different contributions to the time resolution the scintillator cube and the WLS fiber were illuminated with the 355~nm laser light. This light doesn't excite the first scintillator dye. 355~nm wavelength corresponds to the maximum in the absorption spectrum of the second dye and to the tail of the absorption spectrum of the WLS fiber. Therefore, this wavelength excites the second scintillator dye POPOP and with a small efficiency the WLS fiber. Illumination was performed at 7.5~cm distance from the SiPM using 134~cm fiber without mirror at the far end. The results are also shown in figure~\ref{fig:laser_resolution}. The WLS fiber alone gives much better time resolution up to 0.3~ns for 40~p.e. signals. The illumination of the second scintillator dye also results in a better time resolution of 0.5~ns for 40~p.e. signals. These results demonstrate that the contribution from the WLS fiber with about 7.5~ns decay time~\cite{Fibers} does not completely dominate the time resolution. Fluctuations in the light absorption and emission from both scintillator dyes also degrade the time resolution, although their decay times of 1-2~ns~\cite{Lakowicz2006} are much shorter.  

The intrinsic single channel  time resolution obtained using the laser is about 0.62~ns for 56~p.e. signals typical for minimum ionizing particles at the SFGD prototype tests at CERN~\cite{SFGD_CERN_test}.
The difference in quadratures between the time resolution for a single channel
obtained using muon data (0.97~ns)
and the time resolution determined
using the laser source is  about 0.75~ns.
 This value is consistent with the expected contribution from the SuperFGD electronics which is 2.5~ns/$\sqrt{12}$ = 0.72~ns. Such agreement indicates that the time resolution obtained with the SuperFGD prototype using calibration procedures presented in this paper is not far from its ultimate limit.

\section{Conclusions}

The time calibration with the time-walk correction procedure for the SuperFGD prototype was developed. It automatically takes into account different time delays along fibers due to different distances from the scintillator cubes to SiPMs and time shifts in particular channels. After the time calibration the time resolution for a single channel is measured to be 0.97~ns.
It was shown that the time resolution improves with amplitude, approaching 0.87~ns for large pulses.
Averaging signals from two fibers 
for one scintillator cube improves the time resolution to 0.68~ns. Averaging the time from 2 scintillator cubes with 2 fibers each improves the time resolution to 0.47~ns which is much better than the electronics intrinsic time resolution of 0.72~ns in one channel due to the time recording step of 2.5~ns.
Averaging time values recorded by several different channels improves the time resolution like $1/\sqrt{N}$, where N is the number of averaged cubes. This means that the contribution to the time resolution from the electronics digitization step of 2.5~ns is also averaged out.  
Therefore, a very good time resolution should be achievable for neutrons since  recoil protons or other produced particles hit typically several cubes and in addition produce larger amplitudes than muons.
Moreover, for neutrons time signals from three fibers per cube  can be used for averaging while in the present study only X and Y fibers were used.
The time resolution obtained with a 266~nm laser and a fast oscilloscope is about 0.62~ns for signal amplitudes typical for minimum ionizing particles. This intrinsic resolution of the scintillator cube and WLS fiber system, added in quadrature with the contribution from the 2.5 ns digitization of the superFGD electronics, is consistent with the time resolution obtained with the SuperFGD prototype. This shows that the time resolution obtained in this study with the SuperFGD prototype is close to the ultimate limit.

\section*{Acknowledgements}

This work is supported by the Ministry of Science and Higher Education of the Russian Federation (Grant "Neutrino and astroparticle physics" No. 075-15-2020-778); JSPS KAKENHI Grant Numbers JP26247034, JP16H06288, and JP20H00149; SEIDI-MINECO under Grants No.~PID2019-107564GB-I00 and SEV-2016-0588. IFAE is partially funded by CERCA.

\pagebreak

\end{document}